\documentclass[aps,twocolumn,superscriptaddress]{revtex4}

\usepackage{graphicx}
\usepackage{epsfig}             
\usepackage{array}

\begin{document}


\title{A quantum study of multi-bit phase coding for optical storage}

\author{Magnus~T.~L.~Hsu}
\affiliation{ARC COE for Quantum-Atom Optics, Department of Physics, Australian National University, ACT 0200, Australia.}

\author{Vincent~Delaubert}
\affiliation{ARC COE for Quantum-Atom Optics, Department of Physics, Australian National University, ACT 0200, Australia.}
\affiliation{Laboratoire Kastler Brossel, Universit\'e Pierre et Marie Curie, case 74, 75252 Paris cedex 05, France.}

\author{Warwick~P.~Bowen}
\affiliation{Physics Department, University of Otago, Dunedin, New Zealand.}

\author{Claude~Fabre}
\affiliation{Laboratoire Kastler Brossel, Universit\'e Pierre et Marie Curie, case 74, 75252 Paris cedex 05, France.}

\author{Hans-A.~Bachor}
\affiliation{ARC COE for Quantum-Atom Optics, Department of Physics, Australian National University, ACT 0200, Australia.}

\author{Ping~Koy~Lam}
\affiliation{ARC COE for Quantum-Atom Optics, Department of Physics, Australian National University, ACT 0200, Australia.}
\email{Email: ping.lam@anu.edu.au.}

%


\date{\today}

\begin{abstract}
We propose a scheme which encodes information in both the longitudinal and spatial transverse phases of a continuous-wave optical beam. A split detector-based interferometric scheme is then introduced to optimally detect both encoded phase signals. In contrast to present-day optical storage devices, our phase coding scheme has an information storage capacity which scales with the power of the read-out optical beam. We analyse the maximum number of encoding possibilities at the shot noise limit. In addition, we show that using squeezed light, the shot noise limit can be overcome and the number of encoding possibilities increased. We discuss a possible application of our phase coding scheme for increasing the capacities of optical storage devices. 
\end{abstract}

\maketitle





\section{Introduction}



The optical compact disc (CD) was developed in 1979 as a collaboration between two corporations - Sony and Philips.  Today, the CD has wide-ranging storage applications from the audio-visual to computer industries.  The CD system is based on a 780~nm laser (laser spot size of 2.1~$\mu$m) with a storage capacity of approximately 670~MB.  Since their introduction there has been increasing demand for greater storage capacities in optical discs.  The digital versatile disc (DVD), based on a 650~nm laser system (spot size of 1.3~$\mu$m), was introduced. Depending on the format, it can have storage capacities ranging from 4.7~GB to 17.1~GB.  More recently, the HD DVD and Blu-Ray discs based on a 405~nm laser system were released.  The HD DVD system has a spot size of 0.76~$\mu$m and storage capacities of 15~GB to 45~GB, whilst Blu-Ray disc systems have a smaller spot size of 0.6~$\mu$m, with capacities of 25~GB to 100~GB \cite{torok}.

Whilst nano-technology allows the fabrication of mechanical surfaces with nano-meter size structures, it is the diffraction limit of the read-out optics that prevents data storage densities beyond those of present day systems. The trend of compacting more information into an optical disc therefore primarily relies on the use of shorter wavelength lasers to achieve smaller diffraction limited spot sizes. One could envisage that in the not too distant future, such improvement in the storage density would require the use of very short wavelength light beyond the ultra-violet spectrum that cannot be generated with known laser optics.

To date, there have been a number of suggested alternatives for obtaining higher optical storage densities.  For example, currently under development are next generation holographic devices, the holographic versatile disc (HVD), which have capacities exceeding 300~GB through the usage of volume storage.  An alternative approach depends on the encoding of spatial details beyond the diffraction limit of the read-out laser beam \cite{torok}. This approach requires the use of near-field microscopy techniques to resolve sub-diffraction limited features.

In this paper, we revisit a well known alternative of using interferometric techniques \cite{kolobov, sokolov, abouraddy, pittman, bennink, bennink1, gatti1, gatti, kolobov1, lugiato} to extend optical storage density.  We propose to perform multi-bit phase-front coding of optical beams in an interferometric setup.  Our scheme does not require holographic nor near-field optics, instead it utilises two classes of phase coding - the {\it longitudinal} and {\it spatial transverse} phases of an optical beam.  We encode information in the longitudinal phase of a beam, which could take discrete values ranging from 0 to $\pi$. The total number of encode-able phase values scales with the power of the read-out optical beam. We then introduce an extra encoding degree of freedom, the spatial transverse phase-front profile of the beam.  Although not explicitly mentioned in our analysis, it is important to note that the spatial features of the beam used could, in principle, be at the diffraction limit.  In order to resolve the encoded longitudinal phase of the beam, an interferometric scheme is required.  To resolve the spatial phase profile of the beam, a multi-segment photo-detector (e.g. a charge-coupled device (CCD)) can be used. 

This paper is sectioned as follows.  We first reduce our analysis of spatial phase-front beam encoding to the situation of a two-pixel array detector, the split detector \cite{santha, putman, guo, simmons, gittes, denk}. We identify the possible phase profiles symmetric with a split detector and give a modal analysis for these spatial profiles. We also introduce the longitudinal phase of the beam and show how an interferometric scheme based on split detectors can be used to simultaneously obtain information on the longitudinal and spatial phases. A quantum analysis of the measurement noise is then presented. We identify the maximum number of encode-able longitudinal phases at the shot noise limit (SNL). We then show that using squeezed light, one can overcome the shot noise limit and thus the number of encoding possibilities can be further increased. Consequently, we analyse the spectral properties of the signal and noise of the encoded beam. We compare single and consecutive measurement techniques, and show that consecutive measurements provides an improved signal-to-noise ratio (SNR), whilst ensuring compatibility with squeezed light frequency regimes.

\section{Classical Phase Coding}

In this paper, we consider the detection of a specific set of possible
transverse and longitudinal phase transformations of a TEM$_{00}$ field.  To detect such transformations requires a spatially selective detection system
such as a CCD array, split detector or specifically configured
homodyne detector \cite{hsu}.  Split detectors in particular offer
fast response (in the GHz regime) and high efficiency.  These factors
are critical in high bandwidth optical systems.  We therefore
concentrate our analysis on split detectors in this paper.

Restricting our analysis to one-dimension, the set of eigenmodes that
best describe split detectors is the flipped eigenmode basis
\cite{fabre}.  The normalised beam amplitude function $u_{{\rm
f}n}(x)$ for the flipped eigenmode (henceforth termed simply the
{\it flipped mode}) of order $n$ is defined by a TEM$_{pq}$ mode with a
$\pi$ phase flip at the centre of the mode \cite{fabre, delaubert}
\begin{equation}
 u_{{\rm f}n}(x) = \left\{ \begin{array}{ll}
 u_{n}(x) & \textrm{for } x > 0 \\
 -u_{n}(x) & \textrm{for } x < 0.
 \end{array} \right.
 \label{flipp}
\end{equation}
where $u_{n}(x)$ is the one-dimensional normalised beam amplitude
function of a TEM$_{pq}$ mode. 
\begin{figure}[!ht]
    \begin{center}
    \includegraphics[width=7cm]{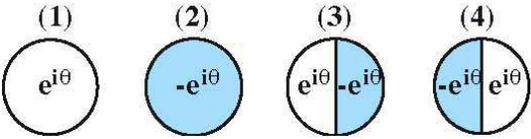}
    \caption{The four possible phase-front profiles resulting from the
    transformation on the input $u_{0}(x)$ beam.}
    \label{flippedmodes}
    \end{center}
\end{figure}

To encode split detector compatible information on the phase-front of
a TEM$_{pq}$ beam, $\pi$ phase flips of this kind should therefore be
applied.  This results in four possible bit values, corresponding to
the four possible two-pixel $\pi$ phase shifts on $u_{0}(x)$, as
illustrated in Fig.~\ref{flippedmodes}. A longitudinal phase factor $e^{i \theta}$ is also introduced to increase the total number of encoding possibilities. The phase coded modes introduced in Fig.~\ref{flippedmodes} are described by the following
transformation
\begin{eqnarray}
u_{0}(x) & \stackrel{(1)}{\longrightarrow} & e^{i \theta}u_{0}(x) \label{1}\\
u_{0}(x) & \stackrel{(2)}{\longrightarrow} & -e^{i \theta} u_{0}(x) \label{2}\\
u_{0}(x) & \stackrel{(3)}{\longrightarrow} &  e^{i \theta}u_{\rm f0}(x) \label{3}\\
u_{0}(x) & \stackrel{(4)}{\longrightarrow} & - e^{i \theta} u_{\rm f0}(x).
\label{4}
\end{eqnarray}

\begin{figure}[!ht]
    \begin{center}
    \includegraphics[width=7cm]{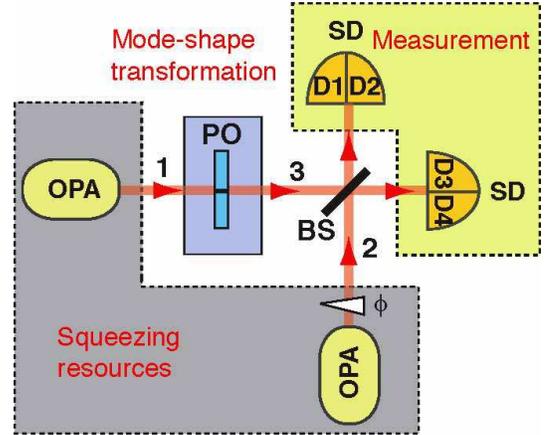}
    \caption{Two-pixel phase coding scheme. BS: beam-splitter, SD: split-detector. D1, D2, D3, D4: labels for the split detector segments. OPA: optical parametric amplifier, PO: phase object.}
    \label{scheme}
    \end{center}
\end{figure}

In order to resolve the four possible phase-front profiles of
Eqs.~(\ref{1})-(\ref{4}), we propose the phase coding scheme shown in
Fig.~\ref{scheme}. Beam 1 undergoes a phase-front
transformation upon traversing a phase object (PO), resulting in transformed beam 3.  Subsequently, beams 2 and 3 are combined on a 50:50 beam-splitter and the two output beams are detected using split detectors.  Each field can be represented by
the positive frequency part of its mean electric field
$\mathcal{E}^{+} e^{-i\omega t}$, where $\omega$ is the optical
frequency.  We are interested in the transverse information described
fully by the slowly varying field envelope $\mathcal{E}^{+}$.  For a
measurement performed in an exposure time $T$, the mean field for
input beams 1 and 2 are given by
\begin{eqnarray}
\mathcal{E}_{1}^{+} (x,t) & = &i \sqrt{\frac{\hbar \omega}{2\epsilon_{0} c T }} \alpha_{0}(t) u_{0}(x) \\
\mathcal{E}_{2}^{+} (x,t) & = & ie^{i\phi} \sqrt{\frac{\hbar
\omega}{2\epsilon_{0} c T }} \beta_{0}(t) u_{0}(x)
\end{eqnarray}
where $\alpha_{0}(t)$ and $\beta_{0}(t)$ are the coherent amplitudes of the
TEM$_{00}$ input beams 1 and 2, respectively. $\phi$ represents the longitudinal phase of beam 2. 

The photo-current signal for all segments of the split detectors are then calculated.  The photo-current signal is related directly to the mean photon number, given by $n_{x<0} = \frac{2\epsilon_{0}cT }{\hbar \omega} \int_{-\infty}^{0} dx
\mathcal{E}^{\dagger} \mathcal{E}$ and $n_{x>0} = \frac{2 \epsilon_{0}
cT}{\hbar \omega} \int_{0}^{\infty} dx \mathcal{E}^{\dagger}
\mathcal{E}$, for split detector $x<0$ and $x>0$ segments, respectively.
\begin{table*}[!htb]
\caption{The photo-current signal terms.}
\label{resultsdc}
\centering
\begin{tabular}{c|c|c|c|c}
 Transformed & Combination {\bf A} & Combination {\bf B} & Combination {\bf C}
& Combination {\bf D} \\
 Mode & $(D_{1} - D_{2}) + (D_{3} - D_{4})$ & $(D_{1} + D_{2}) + (D_{3} +
D_{4})$ & $(D_{1} - D_{2}) - (D_{3} - D_{4})$ & $(D_{1} + D_{2}) -
(D_{3} + D_{4})$ \\
\hline \hline
 $e^{i \theta}u_{0}(x)$ & 0 & $\alpha_{0}^{2} + \beta_{0}^{2}$ & 0 & $2
\alpha_{0} \beta_{0} \sin(\phi-\theta)$ \\
 $-e^{i \theta}u_{0}(x)$ & 0 & $\alpha_{0}^{2} + \beta_{0}^{2}$ & 0 & $-2
\alpha_{0} \beta_{0} \sin(\phi-\theta)$ \\
 $e^{i \theta}u_{{\rm f}0}(x)$ & 0 & $\alpha_{\rm f0}^{2} + \beta_{0}^{2}$ & $2
\alpha_{\rm f0} \beta_{0} \sin(\phi-\theta)$ & 0 \\
$-e^{i \theta}u_{{\rm f}0}(x)$ & 0 & $\alpha_{\rm f0}^{2} + \beta_{0}^{2}$ & $-2 \alpha_{\rm f0} \beta_{0} \sin(\phi-\theta)$ & 0
\end{tabular}
\end{table*}

We consider four possible combinations for the pair-wise photo-current addition and subtraction, with the photo-current signal terms shown in Table~\ref{resultsdc}. The table shows that for both combinations {\bf A} and {\bf B}, all four mode transformations have identical signal values of $0$ and $\alpha_{0}(t)^{2} + \beta_{0}(t)^{2}$, respectively.  Thus the phase-front transformation on the input beam
cannot be determined.  Combinations {\bf C} and {\bf D}, on the other
hand, allow the $u_{{\rm f}0}(x)$ and $u_{0}(x)$ modes to be detected
respectively with a sign change for the different phase-front
transformation. Moreover, the phase coding scheme is sensitive to the differential longitudinal phase $(\theta - \phi)$.

Note that $\phi$ has to be calibrated in order to determine the encoded phase $\theta$. $\phi$ is scanned between 0 and $\pi$ until an extremum value of $\pm 2 \alpha \beta$ is obtained. For example, for combination {\bf D}, a value of $-2 \alpha_{0} \beta_{0}$ for $\phi=\phi_{\rm opt}$ tells us that the encoded mode-shape is $-u_{0}(x)$ with longitudinal phase $\theta = \pi/2 - \phi_{\rm opt}$. The maximal signal is obtained for a phase difference of $\phi - \theta = \pi/2$. This interferometric scheme therefore enables a unique distinction of all of the four phase-front transforms given in Eqs.~(\ref{1})-(\ref{4}).

\section{Quantum Phase Coding}

We would like to quantify the maximum number of encoding possibilities, whose limit is ultimately set by the SNL. The SNL is identified and ways to improve the sensitivity of the measurement using squeezed light are shown. Consider the field operators in the sideband frequency domain, $\Omega$. For brevity, we do not explicitly denote the frequency dependence for the field operators hereon, which are given by
\begin{eqnarray} \label{gen}
\hat{\mathcal{E}}^{+}_{1} & = & i\sqrt{\frac{\hbar \omega}{2
\epsilon_{0}cT}} \left( \alpha_{0}u_{0}(x) + \sum_{n=0}^{\infty}
\delta \hat{a}_{n} u_{n}(x) \right) \\
\hat{\mathcal{E}}_{2}^{+} & = & ie^{i\phi} \sqrt{\frac{\hbar \omega}{2
\epsilon_{0}cT}} \left( \beta_{0}u_{0}(x) + \sum_{n=0}^{\infty}
\delta \hat{b}_{n} u_{n}(x) \right),
\end{eqnarray}
where the first terms are the coherent amplitude of the $u_{0}(x)$
mode. The second terms are the quantum fluctuations $\delta \hat a = \hat a - \langle \hat a \rangle$ and $\delta \hat b = \hat b - \langle \hat b \rangle$, with $\hat a$ and $\hat b$ being annihilation operators, of beams 3 and 2 in Fig.~\ref{scheme}, respectively. Depending on the phase-front transformation on beam 1, the field operator describing beam 3 is given by
\begin{eqnarray}
    \hat{\mathcal{E}}_{3}^{(1)+} & = & ie^{i\theta} 
    \sqrt{\frac{\hbar \omega}{2 \epsilon_{0}cT}} \left( \alpha_{0} u_{0}(x) +
    \sum_{n=0} \delta \hat{a}_{n} u_{n}(x) \right)
\nonumber\\
    \hat{\mathcal{E}}_{3}^{(2)+} & = & - ie^{i\theta} 
    \sqrt{\frac{\hbar \omega}{2 \epsilon_{0}cT}} \left( \alpha_{0} u_{0}(x) +
    \sum_{n=0}  \delta \hat{a}_{n} u_{n}(x) \right)
\nonumber\\
    \hat{\mathcal{E}}_{3}^{(3)+} & = & ie^{i\theta} 
    \sqrt{\frac{\hbar \omega}{2 \epsilon_{0}cT}} \left( \alpha_{\rm f0} u_{\rm f0}(x) +
    \sum_{n=0} \delta \hat{a}_{\rm fn} u_{\rm fn}(x) \right)
\nonumber\\
    \hat{\mathcal{E}}_{3}^{(4)+} & = & - ie^{i\theta} 
    \sqrt{\frac{\hbar \omega}{2 \epsilon_{0}cT}} \left( \alpha_{\rm f0} u_{\rm f0}(x) +
    \sum_{n=0} \delta \hat{a}_{\rm fn} u_{\rm fn}(x) \right)
\nonumber\\
\end{eqnarray}
where the superscript denotes the transformations corresponding to Eqs.~(\ref{1})-(\ref{4}). 

The RF photo-current for all segments of the split detectors are then
calculated similarly to the previous section. The overlap integrals in the expressions for the photo-current sum and difference operators are simplified using the respective orthogonality properties of the $u_{n}(x)$ and $u_{{\rm
f}n}(x)$ modes \cite{treps1d, treps2d, treps2da}.

The photo-current noise corresponding to combinations \textbf{C} and \textbf{D} are shown in Table~\ref{results}, where we have defined the quadrature noise operator as $\delta \hat{X}_{a}^{\psi} = e^{-i \psi} \delta \hat{a} + e^{i \psi} \delta \hat{a}^{\dagger}$. We assume the phase-front transformation is lossless. Therefore  the photon statistics of the transformed field is unchanged relative to the initial field, so that $|\alpha_{\rm f0}| = |\alpha_{\rm 0}| = |\alpha|$ and $\langle ( \delta \hat X_{a_{\rm f0}}^{\psi} )^2 \rangle = \langle ( \delta \hat X_{a_{0}}^{\psi} )^2 \rangle =  \langle ( \delta \hat X_{a}^{\psi} )^2 \rangle$. Table~\ref{results} therefore suggests that squeezing the input beams 1 and 2 will lead to enhanced noise  performances.
\begin{table*}[!htb]
\caption{Photo-current noise.}
\label{results}
\centering
\begin{tabular}{c|c|c}
\hline
 Transformed & Combination {\bf C} & Combination {\bf D} \\
 Mode & $(D_{1} - D_{2}) - (D_{3} - D_{4})$ & $(D_{1} + D_{2}) - (D_{3} + D_{4})$\\
\hline \hline
 $e^{i \theta}u_{0}(x)$ & $-\beta_{0} \delta \hat{X}_{a_{\rm f0}}^{(\phi - \theta + \pi/2)} + \alpha_{0} \delta \hat{X}_{b_{\rm f0}}^{(\theta - \phi + \pi/2)}$ & $- \beta_{0} \delta \hat{X}_{a_{0}}^{(\phi - \theta +\pi/2)} + \alpha_{0} \delta \hat{X}_{b_{0}}^{(\theta - \phi + \pi/2)}$ \\
 $-e^{i \theta}u_{0}(x)$ & $\beta_{0} \delta \hat{X}_{a_{\rm f0}}^{(\phi - \theta + \pi/2)} - \alpha_{0} \delta \hat{X}_{b_{\rm f0}}^{(\theta - \phi + \pi/2)}$  & $ \beta_{0} \delta \hat{X}_{a_{0}}^{(\phi - \theta +\pi/2)} - \alpha_{0} \delta \hat{X}_{b_{0}}^{(\theta -\phi + \pi/2)}$  \\
 $e^{i \theta}u_{{\rm f}0}(x)$ & $- \beta_{0} \delta \hat{X}_{a_{\rm f0}}^{(\phi - \theta +\pi/2)}  + \alpha_{\rm f0} \delta \hat{X}_{b_{0}}^{(\theta - \phi +\pi/2)}$ & $- \beta_{0}
\delta \hat{X}_{a_{0}}^{(\phi - \theta + \pi/2)} + \alpha_{\rm f0} \delta \hat{X}_{b_{\rm f0}}^{(\theta - \phi + \pi/2)}$  \\
 $-e^{i \theta}u_{{\rm f}0}(x)$ & $ \beta_{0} \delta \hat{X}_{a_{\rm f0}}^{(\phi - \theta +\pi/2)}  - \alpha_{\rm f0} \delta \hat{X}_{b_{0}}^{(\theta - \phi +\pi/2)}$ & $\beta_{0}
\delta \hat{X}_{a_{0}}^{(\phi - \theta + \pi/2)} - \alpha_{\rm f0} \delta \hat{X}_{b_{\rm f0}}^{(\theta - \phi + \pi/2)}$ \\
\hline
\end{tabular}
\end{table*}

We now determine the maximum number of encoding possibilities in our phase coding scheme. The encoding limit is determined by the minimum longitudinal phase difference detectable $\Delta \theta_{\rm min}$. This corresponds to when the signal and noise variances are equal (i.e. SNR =1). 

The SNR is calculated by taking the ratio of the signal and noise variances, given by $\langle n \rangle^{2}$ and $\langle \delta \hat{n}^{2} \rangle$, respectively. The corresponding SNR of the measurement for the $\pm u_{0}$ and $\pm u_{\rm f0}$ modes are denoted by ${\cal R}_{\bf C}$ and ${\cal R}_{\bf D}$, respectively, and have the same form given by
\begin{equation} \label{SNRC}
{\cal R} =  \frac{4\alpha^{2} \beta_{0}^{2} \sin^{2}(\phi-\theta)}{ \alpha^{2} \langle (\delta \hat{X}_{b_{0}}^{\psi})^{2} \rangle + \beta^{2} \langle (\delta \hat{X}_{a}^{\psi})^{2} \rangle}
\end{equation}

If the input beams are coherent with $\langle ( \delta \hat X^{\psi}_{a} )^2 \rangle = \langle ( \delta \hat X^{\psi}_{b_{0}} )^2 \rangle = 1$, then ${\cal R}^{\rm coh} = 4\alpha^{2} \beta_{0}^{2} \sin^{2}(\phi-\theta)/ (\alpha^{2} + \beta_{0}^{2})$. This coherent state SNR serves as a benchmark for which to compare the SNR achievable with squeezing.  For quadrature squeezed input beams 1 and 2 (i.e. $\langle (\delta \hat{X}_{a}^{\psi})^{2} \rangle, \langle (\delta \hat{X}_{b}^{\psi})^{2} \rangle < 1$), we see directly that ${\cal R}^{\rm sqz} > {\cal R}^{\rm coh}$. Squeezed input beams therefore increase the SNR achievable for all possible mode transformations.

Note that in the limit $\beta \gg \alpha$, our phase imaging scheme reduces to that of a homodyne measurement with a SNR given by $\mathcal{R}_{\rm hom} = 4 \alpha^{2} \sin^{2}(\phi-\theta)/ \langle (\delta \hat{X}_{a}^{\psi})^{2} \rangle$. The signal and noise contribution arise from the transverse mode defined by the local oscillator mode-shape.

The minimum longitudinal phase difference detectable (i.e. SNR =1) is given by
\begin{equation}
\Delta \theta_{\rm min}= \sin^{-1} \sqrt{\frac{\alpha^{2} \langle ( \delta \hat{X}_{b_{0}} )^{2} \rangle +  \langle ( \delta \hat{X}_{a} )^{2} \rangle \beta_{0}^{2}  }{4\alpha^{2} \beta_{0}^{2}}}
\end{equation}
where we have assumed that the phase difference between beams 2 and 3 have been optimised to $\phi - \theta = \pi/2$. Since the longitudinal phase of beam 3 is determined modulo $\pi$, the total number of resolvable phase levels is $\pi/\Delta \theta_{\rm min}$, which scales with the power of the read-out optical beam $\alpha$. Note that this contrasts with conventional optical storage devices which are restricted to only two encode-able values (i.e. `0' and `1'), with a SNR proportional to the power of the read-out beam.

Including the four possible transverse encoding combinations, the total number of encode-able levels for our phase coding scheme is therefore given by
\begin{equation}
L_{\rm max}  = \frac{4 \pi}{\Delta \theta_{\rm min}}
\end{equation}

We now consider $L_{\rm max}$ levels for a fixed optical power and show how this can be improved via squeezing. The mean number of photons per bandwidth-time in an optical field $\bar{n}$ can be related to its coherent amplitude $\alpha$, amplitude $\langle (\delta \hat X^+)^2 \rangle$ and phase $\langle (\delta \hat X^- )^2 \rangle$ quadrature noise variances by
\begin{equation}
\bar{n} = \frac{1}{4} (\alpha^2 + \langle (\delta \hat X^+)^2
\rangle + \langle (\delta \hat X^- )^2 \rangle -
2).\label{photonnumber}
\end{equation}
The first thing to note is that for squeezed states, $\bar{n}$ is
non-zero even when $\alpha=0$.  Indeed, as the squeezing increases
(for amplitude squeezing $\langle (\delta \hat X^+)^2
\rangle\rightarrow 0$ and $ \langle (\delta \hat X^- )^2 \rangle
\rightarrow \infty$) $\bar{n}$ increases monotonically. With regards to our
phase-front detection scheme, these photons do not contribute to
the signal, $\alpha$, and therefore for a given optical power, $\bar{n}$, a balance must be obtained between using photons to minimise the noise (maximise the squeezing), and to maximise the signal (maximise $\alpha$).

\begin{figure}[!ht]
    \begin{center}
    \includegraphics[width=7cm]{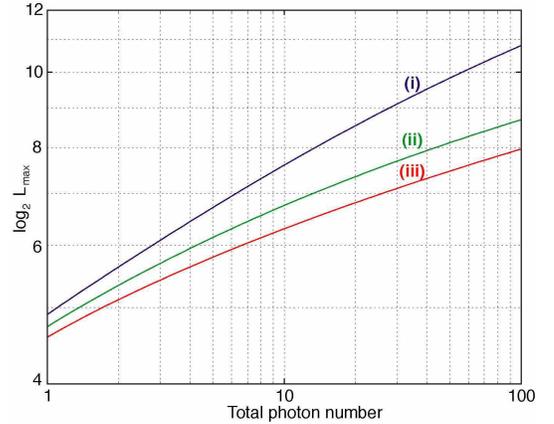}
    \caption{Maximum $\log_{2} L_{\rm max}$ levels of the phase-front coding scheme as a function of the total mean number of photons / (Hz.s) utilised in the protocol. (i) Both input beams amplitude squeezed, (ii) beam 1 amplitude squeezed and beam 2 coherent, (iii) both input beams coherent.}
    \label{SNRoptimised}
    \end{center}
\end{figure}

In this paper, $L_{\rm max}$ levels for a fixed optical power was maximised numerically. This was performed over all possible ratios of photons used to minimise the noise, and those used to maximise the signal. The total optical power is the sum of the number of photons in each of the input beams, each individually given by
Eq.~(\ref{photonnumber}). We considered three cases.  In the
first case, to provide a benchmark, we considered $L_{\rm max}$ levels when no squeezing was available, and both input beams were coherent. In the second case, squeezing was allowed for beam 1 but beam 2 was coherent. In the
third case, both input beams were squeezed. The maximum $L_{\rm max}$ levels for each of these cases is shown in Fig.~\ref{SNRoptimised}. We see that by far, the best $L_{\rm max}$ levels is achieved when both beams are allowed to be squeezed, with $\sim 25\%$ capacity improvement over the coherent state case when 100 photons per bandwidth-time are used. In the case of only one squeezed beam, the capacity improvement is $\sim 10\%$.

\begin{figure}[!ht]
    \begin{center}
    \includegraphics[width=7cm]{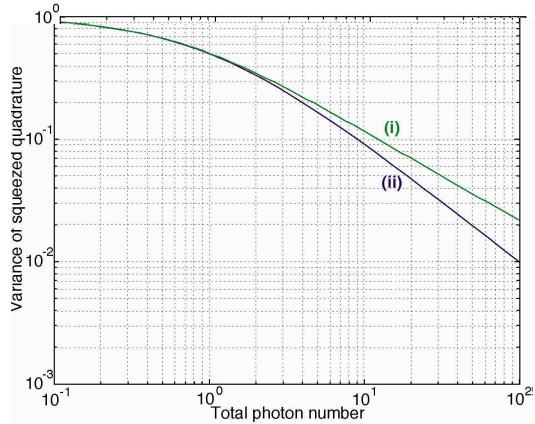}
    \caption{Shot noise normalised level of squeezing required to achieve the optimum number of encode-able levels as a function of the total mean number of photons per bandwidth-time utilised in the protocol. (i) Beam 1 amplitude squeezed and beam 2 coherent, (ii) both input beams amplitude squeezed.}
    \label{VarSQZ}
   \end{center}
\end{figure}

Of course, due to decoherence and inefficiencies, arbitrary levels
of squeezing are not available.  Therefore, it is interesting to
consider not only the maximum $L_{\rm max}$ levels that can be
achieved, but also whether the amount of squeezing required to
achieve it are feasible. The amount of squeezing required to achieve the maximum $L_{\rm max}$ levels for a given total photon number are shown in Fig.~\ref{VarSQZ} for both the one squeezed beam and two squeezed beam cases. In both cases, when
less than 10 photons per time are utilised, squeezing levels below
10~dB are required. Although challenging, such levels of squeezing
are experimentally achievable.  For more than 10 photons per
bandwidth-time however, the level of squeezing required to
achieve the maximum encode-able levels fast becomes unfeasible.
Therefore, utilising squeezed light in the phase-front detection
scheme presented here will only be beneficial when less than
10 photons are available per measurement interval.  

Using coherent beams gives a large number of encode-able levels. For example, assuming idealised shot noise limited performance, we can encode a maximum of $2^{20}$ levels, for a 1~mW beam ($\lambda = 1\mu$m) in the limit that $\beta_{0} \gg \alpha$, during a 1~$\mu$s measurement time (assuming ${\rm SNR} = 1$). Squeezing can further improve the maximum encode-able levels in the limit of low laser power. Similar to the multi-bit encoding schemes of P.~T\"or\"ok's group \cite{torok}, our scheme is a significant improvement over current technology where only one bit per site is encoded.

\section{Potential Application to Optical Storage}

In this section, we investigate the compatibility of our phase coding scheme with an optical disc read-out scheme.

\subsection{Optical disc scheme}

In conventional CDs, the information is encoded in binary format,
by burning physical indentations (commonly termed `pits') on the
disc. Regions where no physical indentations exist are termed
`lands'. A transition between `pit' and `land' encodes for `1', whereas no transition encodes for `0' \cite{williams}. The reflected beam intensity from a focussed laser beam onto the disc allows bit read-out, as the beam undergoes large diffractive losses when impinging on a `land'-`pit' transition.

\begin{figure}[!ht]
    \begin{center}
    \includegraphics[width=7cm]{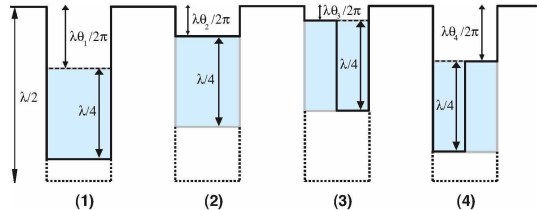}
    \caption{Examples of bit encoding allowing denser information storage on optical discs. The binary information is encoded in the transverse and longitudinal phases of the reflected read-out laser beam. The depth of the `pit' can range from discrete values between 0 and $\lambda/2$. Hence $\theta_{1} \in ( \lambda/4 , \lambda/2 ]$ and $\theta_{2} \in ( 0, \lambda/4 ]$, where  $\lambda$ is the laser wavelength and $\theta$ is the longitudinal phase shift of the laser beam. Note that the `pit' depth is half that of the phase shift experienced by the read-out laser beam due to a round-trip propagation from the optical disc.}
    \label{CD}
   \end{center}
\end{figure}

We propose to store more than one level on a single `pit' site,
by encoding levels as shown on Fig.~\ref{CD}. The information is
contained in the longitudinal and transverse phase domains for each
`pit'. The longitudinal phase is determined by the depth of the `pit', $\lambda \theta / (2 \pi)$, while the transverse phase profile is determined from its shape. Identical beam profiles to those of Fig.~\ref{flippedmodes} can thus be generated. The reflection of the read-out beam has one of the four possible transverse mode profiles, $\rm \pm u_{0}(x)$ or $\pm u_{\rm f0}(x)$, with an additional global phase shift $\theta$. This reflected beam (i.e. beam 3) can then be combined with beam 2 on a 50:50 beam-splitter, as shown in Fig.~\ref{scheme} to perform the phase information decoding. 

Thus far, our analysis has not considered the spectral properties of the read-out signal and noise. Since the optical disc is spinning and the laser read-out time is limited, the spectral power density of a realistic optical disc detection differs from that of an idealised static phase sensing scheme. We examine these issues in the following subsections.

\subsection{Spectral power density for a single measurement}

\begin{figure}[!htbp]
\begin{center}
\includegraphics[width=7cm]{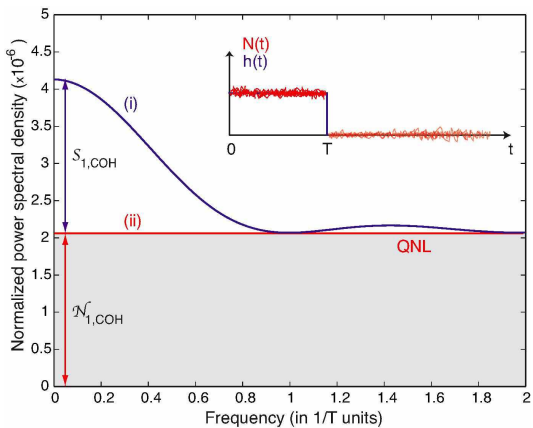}
\caption{The normalised (i) signal $\mathcal{S}_{1}(\nu)$ and (ii) noise $\mathcal{N}_{1}(\nu)$ PSD for a single measurement in a time interval $T$. $\mathcal{S}_{1}(0)$ and $\mathcal{N}_{1}(0)$ are the maximum signal and noise powers at DC. The inset shows the sequence corresponding to a single top hat measurement window, $h(t)$, with $N(t)$ photons.} 
\label{spectrumSingle}
\end{center}
\end{figure}

We first consider the information extraction from a time-limited static disc read-out. $N$ photons are detected in a time interval $T$, as represented on the inset of Fig.~\ref{spectrumSingle}, where the integrated photo-current provides the encoded information.

Using a single top hat function as a time measurement window, the Wiener-Khinchine relation yields the signal power spectral density (PSD), $\mathcal{S}_{1}(\nu)$. In the case of a double sided power spectrum \cite{Couch}, the signal PSD is given by
\begin{eqnarray} \label{s1}
\mathcal{S}_{1}(\nu)=T\left[\int_{-\infty}^{\infty}s(\nu'). {\rm sinc}(\pi
T(\nu'-\nu))d\nu'\right]^{2}
\end{eqnarray}
where $\nu$ is the frequency and $s(\nu)$ is the signal linear spectral
density in the limit of an infinite time measurement, defined as $s(\nu)=\sqrt{\mathcal{S}(\nu)}$. $\mathcal{S}(\nu)$ is the
number of photons per bandwidth-time.

For a single measurement,  $s(\nu)$ is given by $s(\nu)~=~N\delta(\nu)$, where $N$ is the number of photons per time in the signal and $\delta(\nu)$ is a delta function centred at DC. Thus Eq.~(\ref{s1}) becomes
\begin{equation}
\mathcal{S}_{1}(\nu)=  N^{2}T {\rm sinc}^{2} (\pi T \nu)
\end{equation}
where the signal PSD has a squared cardinal sine distribution with a maximum at DC. Fig.~\ref{spectrumSingle} shows the normalised signal and noise PSD.

The noise power spectral density, $\mathcal{N}(\nu)=\xi(\nu)^{2}$ is now obtained. We assume that read-out lasers are shot noise limited. Thus the noise linear spectral density is white and proportional to $\sqrt{N}$, given by $\xi(\nu)  = \sqrt{N}$. The integration time in our measurement is $T$ and the noise PSD is given by 
\begin{eqnarray}
\mathcal{N}_{1}(\nu)= NT\left[\int_{-\infty}^{\infty} {\rm sinc}(\pi
T(\nu'-\nu))d\nu'\right]^{2}=\frac{N}{T}
\end{eqnarray}
where the white noise spectrum has an amplitude of $\sqrt{N/T}$. We have chosen $N=1/T^{2}$ for Fig.~\ref{spectrumSingle} such that the noise power is approximately equal to the maximum signal power. 

The typical measurement time of a DVD device is $T=0.1~\mu$s, corresponding to $N=10^{14}$ photons/s for a read-out laser of 20~$\mu$W power at a wavelength of $\lambda=1~\mu$m. The signal spectrum therefore is in the DC to 10~MHz regime. Improvements on the measurement sensitivity beyond the shot noise limit thus requires broadband squeezing from DC to 10~MHz. Although low frequency squeezing has recently been demonstrated \cite{Kirk, Julien}, many technical challenges exist.

An alternative solution, compatible with current technology, could be the sampling or modulation of the read-out beam to artificially shift the signal to a higher frequency range. For optical discs rotating at approximately 10~bit/$\mu$s (4.32~Mb/s for CDs, and 26.16~Mb/s for DVDs), the sampling or modulation frequencies can be at least 1~GHz, which is compatible with squeezing frequency ranges. However, the disadvantage of such an approach is that the photon number in the signal sidebands is low. The majority of the photons are still distributed in the frequency regime around DC. Thus the improvement in the SNR may not be significant.

\subsection{Spectral power density for consecutive measurements}

\begin{figure}[!htbp]
\begin{center}
\includegraphics[width=7cm]{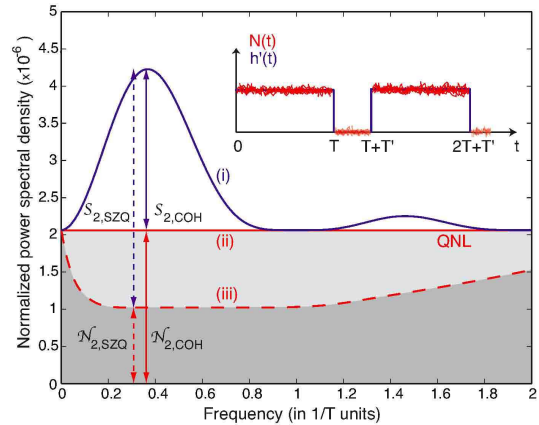}
\caption{(i) Normalised signal $\mathcal{S}_{2}(\nu)$, (ii) shot noise and (iii) squeezed noise $\mathcal{N}_{2}(\nu)$ PSD for two consecutive measurements of time intervals $T$ separated by a delay $T'=0$. The maximum signal and noise powers are respectively, $\mathcal{S}_{\rm 2,COH}$ and $\mathcal{N}_{\rm 2,COH}$, for a coherent state read-out laser at $\nu \sim 0.35/T$. In the case of a squeezed read-out laser, the maximum signal and noise powers are denoted by $\mathcal{S}_{\rm 2,SQZ}$ and $\mathcal{N}_{\rm 2,SQZ}$, respectively. The inset shows the sequence corresponding to two consecutive measurements.}
\label{spectrumConsecutive}
\end{center}
\end{figure}

We now propose to perform consecutive `pit' measurements where the centre of the signal PSD is shifted to a higher frequency. Two consecutive measurements of the variable $N(t)$ during two time intervals of length $T$, separated by a delay $T'$, is shown schematically on the inset of Fig.~\ref{spectrumConsecutive}.

The difference between two consecutive measurements yields a signal PSD given by the contribution of each  individual sine-wave at frequency $\nu$, to the total signal, and averaging over all possible initial phases $\Theta$, giving
\begin{equation}
\mathcal{S}_{2}(\nu)=\mathcal{S}(\nu)\eta_{2}(\nu)
\end{equation}
where $\eta_{2}(\nu)$ is given by
\begin{eqnarray}
\eta_{2}(\nu)=\kappa\Big{\langle}\left(\int_{0}^{T}-\int_{T+T'}^{2T+T'}\sin(2\pi\nu
t+\Theta)dt \right)^{2}\Big{\rangle} \nonumber
\end{eqnarray}
and where $\kappa$ is a normalisation constant. Note that a similar calculation can be applied to the single measurement case.

The signal and noise PSD for $T'=0$ are shown in Fig.~\ref{spectrumConsecutive}. The signal PSD is shifted to the MHz frequency regime (which is more compatible with routinely obtained experimental squeezing frequency regimes). The signal PSD maximum is at $\nu \sim 0.35/T$ for $T'=0$. This is also the regime where the bandwidth is maximum. Increasing $T'$ shifts the maximum signal power to lower frequencies and sharpens the distribution. Thus the bandwidth reduces with increasing $T'$. For $T'=T$, the maximum signal power is for example $\sim 1.5$ times larger than that for the $T'=0$ case and occurs at $\nu \sim 0.2/T$, whereas the bandwidth reduces by half. $T'$ can thus be tuned to obtain an optimum for signal power, bandwidth and compatibility with squeezing frequencies.  

The SNR of interest in our optical memory scheme corresponds to that of a single frequency $\nu$, as defined in Eq.~(\ref{SNRC}). Therefore our proposed optical memory scheme will require a frequency mixer or bandpass filter centred at $\nu$, where the measurement SNR is maximum. 

Differential consecutive measurements is a technique already employed in current optical disc devices, as it allows the cancellation of common-mode classical noise, provided that the phase of the read-out laser is well calibrated in a `pit'-to-`pit' measurement. Furthermore, the maximum of the normalised PSD for consecutive measurements is slightly larger than that for the single measurement case, assuming a coherent state read-out laser with the same parameters. If a broadband 3~dB squeezed state is used as the read-out laser, the SNR doubles for consecutive measurements. However the SNR improvement for the case of a single measurement is negligible. This is because low frequency noise sources (e.g. acoustic noise) overwhelm the squeezing.

\section{Conclusion}

We have presented a scheme to perform longitudinal and transverse spatial phase coding of continuous-wave optical beams. We have shown that by performing selective combinations of photo-current addition and subtraction, the phase coded signal can be extracted. In order to optimise the phase signal, the longitudinal phase of beam 2 has to be calibrated and optimised such that $\phi = \pi/2 + \theta$. Whilst current CD technologies are limited by a number of different noise sources, such as thermal-Johnson noise and electronic noise, our analysis assumes shot noise limited performance. The maximum number of encoding possibilities for this regime was calculated, suggesting significant improvement with our phase coding scheme. 
However, by using squeezed light, the shot noise limit can be overcome and thus the maximum number of encode-able levels increased. We then presented a possible application of our phase coding scheme in increasing the capacities of optical storage devices. We analysed the performance of single measurement techniques and showed that the signal and noise PSD are centred around DC sideband frequencies. We then analysed the PSD of differential consecutive measurements and showed that the PSD spectrum is shifted to higher sideband frequencies. In order to extract the phase signal, frequency mixing or narrow bandpass filtering techniques can be used. The differential consecutive measurement technique provides a good SNR whilst ensuring compatibility with squeezing frequencies. 

Our phase coding scheme can be extended to implement a multi-pixel array detector. Delaubert {\it et al.} \cite{vincent-cd} has performed a quantum study of multi-pixel array detection and shown that it is possible to perform multi-pixel transverse spatial phase encoding. Possible implementation of a multi-pixel scheme would require the incorporation of multiple interferometers and the use of multi-squeezed beams or a multi-mode OPO system \cite{lopez}.

\section*{Acknowledgment}
We would like to thank N.~Treps and C.~C.~Harb for discussions. This work was funded under the Australian Research Council Centre of Excellence and CNRS France. W.~P.~Bowen acknowledges funding from the MacDiarmid Institute for Advanced Materials and Nanotechnology.

\end{document}